\documentclass[trackchanges]{aastex631}
\usepackage{amsmath}
\usepackage{multirow}

\begin{document}

\title{Proton and Alpha Driven Instabilities in an Ion Cyclotron Wave Event}

\newcommand{\ucb}{\affiliation{Physics Department, University of California, Berkeley, CA 94720-7300, USA}}
\newcommand{\ssl}{\affiliation{Space Sciences Laboratory, University of California, Berkeley, CA 94720-7450, USA}}

\author[0000-0001-6077-4145]{Michael D. McManus}
\ssl
\correspondingauthor{Michael D. McManus}
\email{mdmcmanus@berkeley.edu}
\author[0000-0001-6038-1923]{Kristopher G. Klein}
\affil{University of Arizona, Tucson, AZ, USA}
\author[0000-0001-5030-6030]{Davin Larson}\ssl
\author[0000-0002-1989-3596]{Stuart D. Bale}\ucb\ssl
\author[0000-0002-4625-3332]{Trevor A. Bowen}\ssl
\author[0000-0002-9954-4707]{Jia Huang}\ssl
\author[0000-0002-0396-0547]{Roberto Livi}\ssl
\author[0000-0003-0519-6498]{Ali Rahmati}\ssl
\author[0000-0002-4559-2199]{Orlando Romeo}\ssl
\author[0000-0003-1138-652X]{Jaye Verniero}
\affil{Heliophysics Science Division, NASA, Goddard Space Flight Center, Greenbelt, MD 20771, USA}
\author[0000-0002-7287-5098]{Phyllis Whittlesey}\ssl

\begin{abstract}
Ion scale wave events or “wave storms” in the solar wind are characterised by enhancements in magnetic field fluctuations as well as coherent magnetic field polarisation signatures at or around the local ion cyclotron frequencies. In this paper we study in detail one such wave event from Parker Solar Probe’s (PSP) fourth encounter, consisting of an initial period of left-handed (LH) polarisation abruptly transitioning to a strong period of right-handed (RH) polarisation, accompanied by clear core-beam structure in both the alpha and proton velocity distribution functions. A linear stability analysis shows that the LH polarised waves are anti-Sunward propagating Alfvén/ion cyclotron (A/IC) waves primarily driven by a proton cyclotron instability in the proton core population, and the RH polarised waves are anti-Sunward propagating fast magnetosonic/whistler (FM/W) waves driven by a firehose-like instability in the secondary alpha beam population. The abrupt transition from LH to RH is caused by a drop in the proton core temperature anisotropy.
We find very good agreement between the frequencies and polarisations of the unstable wave modes as predicted by linear theory and those observed in the magnetic field spectra. Given the ubiquity of ion scale wave signatures observed by PSP, this work gives insight into which exact instabilities may be active and mediating energy transfer in wave-particle interactions in the inner heliosphere, as well as highlighting the role a secondary alpha population may play as a rarely considered source of free energy available for producing wave activity.

\end{abstract}

\section{Introduction} \label{sec:intro}
Ion velocity distribution functions (VDFs) in the fast-like solar wind are not in local thermodynamic equilibrium (LTE), as evidenced by the presence of temperature anisotropies $T_\perp \neq T_\parallel$ ($T_\perp$ and $T_\parallel$ the perpendicular and parallel temperatures respectively), drifts between different particle populations, and temperature disequilibria between species. These departures from LTE represent sources of free energy that are available to drive kinetic micro-instabilities and produce a variety of ion-scale plasma waves. In turn the particles are scattered by the presence of these waves, driving them back towards LTE and limiting the free energy available to drive the instability. This should be contrasted with the relatively ineffective regulating effect of Coulomb collisions on fast wind ion VDFs, where the Coulomb collisional age $A_c \lesssim 1$ at 1AU \citep{marsch2006kinetic,alterman2018comparison}. Slow wind on the other hand is more likely to have had sufficient time to become collisionally old, $A_c \gg 1$, resulting in the relaxation of several of these non-thermal features (in particular, thermalisation between the proton and alpha temperatures and a smaller alpha-proton drift speed \citep{maruca2013collisional, kasper2008hot,alterman2018comparison}). Plasma waves driven by these departures from LTE are directly measurable in magnetic and electric field data and often probed via examination of magnetic field power spectra, magnetic field ellipticity (polarisation), Poynting vectors, etc \citep{he2011possible,podesta2011magnetic,bowen2020electromagnetic}. Empirically \citep{jian2009ion,jian2010observations,jian2014electromagnetic,gary2016ion,wicks2016proton,klein2018majority,verniero2020parker,bowen2020ion,martinovic2021ion}, it suffices to restrict attention to quasi-parallel propagation with $\mathbf{k}\times\mathbf{B} = 0$, in which case there are three main types of ion instability driven modes physically relevant in the solar wind. Alfvén/Ion Cyclotron (A/IC) instabilities are triggered by temperature anisotropies with $T_\perp/T_\parallel > 1$ and produce left-handed (LH) circularly polarised waves in the particles' rest frame. Fast magnetosonic/whistler (FM/W) instabilities produce right-handed (RH) circularly polarised waves and are driven by temperature anisotropies with $T_\perp/T_\parallel < 1$. Finally ion/ion component instabilities are caused by relative drifts between particle populations (see \cite{verscharen2019multi} for a more comprehensive summary of relevant microinstabilities in the solar wind).

Traditionally, statistical analyses of the role that temperature anisotropy driven instabilities play in the solar wind have focussed on parameterising instability thresholds as functions of the temperature anisotropy $R = T_\perp/T_\parallel$ and parallel plasma beta $\beta_\parallel = 8 \pi n T_\parallel/B^2$ only, where $n$ is the plasma 
 number density and $B$ the magnetic field magnitude. Data plotted as functions of these two parameters are colloquially referred to as ``Brazil plots”. Given the large number of free parameters available when modelling solar wind plasma (each bi-Maxwellian comes with 6), flattening the data onto a 2D parameter space may seem overly restrictive. However, the excellent agreement between the distribution of proton measurements in the $(R_p,\beta_{p,\parallel})$ plane and instability contours of constant growth rate \citep{kasper2002wind,gary2003consequences,hellinger2006solar,bale2009magnetic,matteini2012ion}, as well as statistical enhancements in the amplitude of magnetic field fluctuations at the boundaries of these measurements \citep{bale2009magnetic}, is strong evidence that various kinetic instabilities are indeed active in the solar wind and are limiting the range of proton temperature anisotropies. However, the fact that it is the oblique rather than parallel instabilities that appear to constrain the data in $(R_p,\beta_{p,\parallel})$ space better, while evidence from both statistical surveys and case studies \citep{klein2018majority,klein2019linear,martinovic2021ion,jian2010observations,gary2016ion} suggest that parallel instabilities should be dominant, is perhaps evidence of the shortcomings of this approach and that other effects such as additional species or drift speeds need to be taken into account. Similar effects on the distribution of alpha particle measurements have also been observed \citep{gary2003consequences,maruca2012instability}, suggesting analogous mechanisms at work. A second aspect regarding the effectiveness of linear theory in constraining observations is raised when comparing linear vs non-linear timescales. Previous studies have found non-linear turbulent timescales to be comparable to 
 or faster than linear ones \citep{klein2019linear,bandyopadhyay2022interplay}, suggesting that the assumptions in linear theory of a uniform, homogeneous background (so that a Fourier transform can be performed) might break down due to turbulence. However, recent work \citep{bandyopadhyay2022interplay} has shown that this timescale ordering is reversed near the instability thresholds, causing kinetic instabilities to dominate precisely near the bounds where we observe the solar wind measurements to be constrained. 

In this work we study one such ion-scale wave event in detail, using particle data from the SPAN-I instrument \citep{livi2021span} on board Parker Solar Probe (PSP; \cite{fox2016solar}) to model core and beam components of both the proton and alpha populations, and make detailed comparisons between the predicted properties of unstable wave modes and electromagnetic field observables.

\section{Data} \label{sec:data}
\begin{figure}
    \centering
    \includegraphics[width=0.85\textwidth]{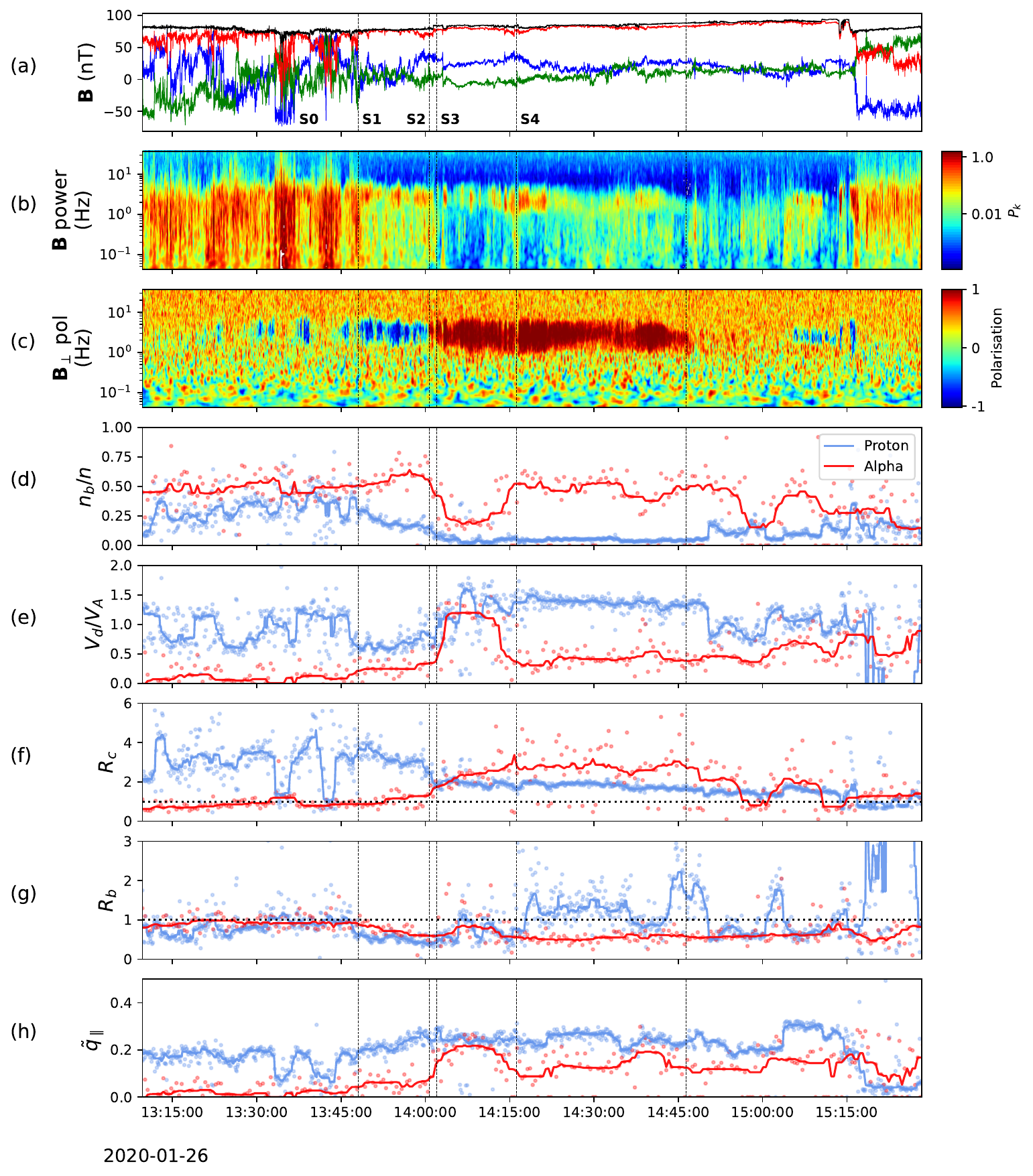 }
    \caption{Ion scale wave event interval showing a period of coherent wave power and polarisation coincident with both proton and alpha particle beam populations. Panels shown are (a) magnetic field $\mathbf{B}$ in spacecraft coordinates, (b) magnetic field trace power spectrum, (c) magnetic field perpendicular polarisation, (d) proton (blue) and alpha (red) beam density ratio $n_b/n$, where $n = n_c + n_b$ for each species, (e) beam drift speed relative to core as a fraction of the local Alfvén speed $V_d/V_A$ for each species, (f) core temperature anisotropy $R_c = T_{c\perp}/T_{c\parallel}$, (g) beam temperature anisotropy $R_b = T_{b\perp}/T_{b\parallel}$, and (h) normalised parallel heat flux $\tilde{q}$. Solid lines in the bottom four panels are moving window medians (15 measurement points wide).}
    \label{fig:event_timeseries}
\end{figure}
The interval studied here is from the inbound portion of PSP's fourth encounter, from 2020-01-26/13:10:00 to 2020-01-26/15:30:00, and is shown in figure \ref{fig:event_timeseries}. The wave event itself that we perform an instability analysis on is demarcated by the leftmost and rightmost dotted lines, from 13:48 to 14:46. High resolution magnetic field measurements from the FIELDS magnetometer \citep{bale2016fields} are shown in panel (a) in spacecraft coordinates, where the $x$ component (blue) points in the spacecraft ram direction, the $z$ component (red) points in the sunward direction, and the $y$ component (green) completes the right-handed set (see the appendix figure in \cite{woodham2021enhanced}). A Morlet wavelet transform \citep{torrence1998practical} is applied to these measurements to compute the magnetic field power spectrum as a function of frequency and time, shown in panel (b), normalised by a Kolmogorov $f^{-5/3}$ power spectrum in order to highlight any coherent wave power above background. Panel (c) plots the perpendicular polarisation of the magnetic field with left-handed (LH) polarisation in blue and right-handed (RH) in red. The power in the left- and right-handed components is computed as
\begin{align}
    P_L(t,f) &= |B_x(t,f) - i B_y(t,f)|^2 \\
    P_R(t,f) &= |B_x(t,f) + i B_y(t,f)|^2,
\end{align}
where $B_i(t,f)$ are the $i$-th components of the wavelet transform of $\mathbf{B}$, in local field-aligned coordinates, from which the perpendicular polarisation is defined as the normalised difference
\begin{equation}
    \text{Pol}(t,f) = \frac{P_R(t,f) - P_L(t,f)}{P_R(t,f) + P_L(t,f)}.
\end{equation}

\section{Fits} \label{sec:fits}
Particle data was obtained from measurements made by the SPAN-I electrostatic analyser. The MPFIT IDL library \citep{markwardt2009non} was used to fit a bi-Maxwellian distribution to both the proton core and beam populations, at their native cadence of 7s, and the fit was carried out in the SPAN-Ion instrument frame. To the alpha SF01 channel, we also fit a bi-Maxwellian distribution to the alpha core and beam, with one additional parameter $\epsilon$ to capture the small amount of protons leaking into the SF01 channel (see \cite{mcmanus2022density}). In order to get slightly better statistics, two SF01 spectra were summed together per fit. From MPFIT, the fit uncertainties for the alpha parameters are approximately 18\% for the densities, 7\% for the core velocity components, 30\% for the drift velocity, 9\% for the beam velocity components, 10\% for the perpendicular temperatures, 20\% for the parallel temperatures, and 20\% for the temperature anisotropies. For the purposes of figure \ref{fig:event_timeseries}, the plasma density $N_e$ used to compute the Alfvén speed $V_A = B/\sqrt{4\pi m_p n_e}$ is the electron density computed via extraction of the plasma frequency line from FIELDS RFS spectra \citep{romeo2021characterization}. This makes the reasonable approximation that $n_p \approx n_e$, as the alpha abundance is quite low, around $1-3\%$ throughout this interval \citep{woolley2021plasma}. 


In the bottom panel of figure \ref{fig:event_timeseries} we compute from the fits the parallel normalised heat flux. For a single species VDF consisting of core and beam bi-Maxwellians, the integral $Q_{\parallel} = \int \frac{1}{2} m  v^2 v_{\parallel} f(\mathbf{v}) d^3\mathbf{v}$ evaluated in that species' centre of mass frame gives
\begin{equation}
        Q_{\parallel} = \frac{1}{2}m \frac{n_c n_b}{n_c + n_b} v_d \biggl( \frac{3}{2} \left(v_{tb\parallel}^2 - v_{tc\parallel}^2\right) +
    v_d^2 \frac{n_c^2 - n_b^2}{(n_c + n_b)^2} + v_{tb\perp}^2 - v_{tc\perp}^2\biggr),
\end{equation}
where $n_c$ ($n_b$) is the core (beam) density, $v_d$ the core-beam drift speed, $v_{tc\perp}$ ($v_{tb\perp}$) the core (beam) perpendicular thermal velocity, and $v_{tc\parallel}$ ($v_{tb\parallel}$) the core (beam) parallel thermal velocity. Thermal velocity is defined as $v_t = \sqrt{2T/m}$, and the centre of mass frame is defined as that in which $n_c \mathbf{V}_c + n_b \mathbf{V}_b = 0$, with  $\mathbf{V}_c,\mathbf{V}_b$ the core and beam velocities respectively. $Q_{\parallel}$ is then normalised by the total density and total parallel temperature $T_\parallel$,
\begin{equation}
    \tilde{q}_{\parallel} = \frac{Q_{\parallel}}{m (n_c + n_b) T_\parallel^{3/2}}.
\end{equation}
$\tilde{q}_\parallel$ is a measure of VDF asymmetry along the magnetic field and can therefore be used to quantify the prominence of a secondary beam. This is particularly useful in the case of the alphas, where the VDFs are often fairly isotropic and automated fitting routines can result in two smaller bi-Maxwellians being fit ``under" one larger one, with roughly equal core and beam densities $n_c \approx n_b$, and very low core-beam drift speeds. An example of this can be seen in figure \ref{fig:event_timeseries} before the wave event starts (which we have labelled S0, defined in the next section), where $n_{\alpha b}/n \approx 0.5$ but $\tilde{q}_{\alpha,\parallel}$ is very small, indicating the absence of a beam and a more isotropic distribution function. 


\section{Event Interval} \label{sec:event_interval}
The wave event is characterised by a strong coherent band of LH magnetic field polarisation that very abruptly (over a time of roughly 1 minute) switches to RH polarisation, and occurs during one of the quiet intervals between two switchback patches \citep{bale2021solar}. From considering this change in polarisation signature, and the results of the bi-Maxwellian fits plotted in figure \ref{fig:event_timeseries}, we split the wave interval into 4 sub-intervals, denoted S1 to S4, and use S0 to denote the period before the wave event starts; these are indicated in figure \ref{fig:event_timeseries} by vertical dotted lines. Figure \ref{fig:example_vdfs} shows example proton and alpha VDFs from S0, S1, S3 and S4. We note that even within each sub-interval there is substantial variation in the VDF parameters (as can be seen in figure \ref{fig:event_timeseries}), however we have attempted to choose examples that reflect the broad stroke qualitative changes. The top row of figure \ref{fig:example_vdfs} shows example proton and alpha VDFs from S0, illustrating the relatively isotropic alpha VDF and less prominent proton beam population prior to the wave event and the appearance of the magnetic field polarisation signature. A brief description of each sub-interval is as follows: 
\begin{figure}
    \centering
    \includegraphics[width=0.9\textwidth]{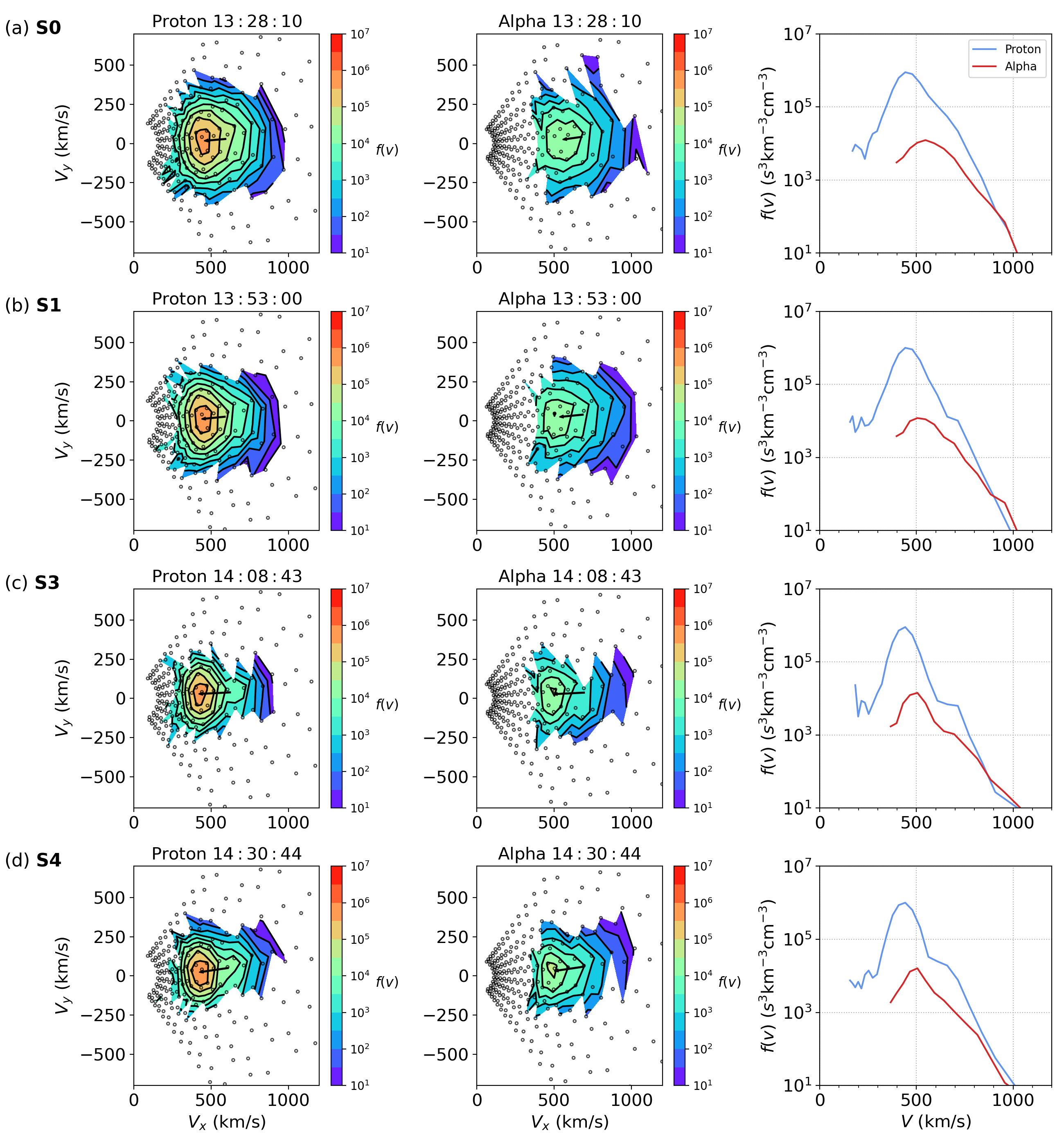}
    \caption{Example proton and alpha VDFs, as well as 1D profiles (summed over all look directions) of $f(v)$ of the proton and alpha spectra at four different times. First time (row (a)) is from before the wave event begins at 13:48:05, with no LH or RH polarisation. Next three rows are from sub-intervals S1, S3 and S4 respectively (see text).}
    \label{fig:example_vdfs}
\end{figure}
\begin{enumerate}
    \item \textbf{S1}: 2020-01-26/13:48:05 to 2020-01-26/14:00:45, period of continuous LH magnetic field polarisation. The second row in figure \ref{fig:example_vdfs} shows example VDFs from S1. Proton VDFs have a relatively  dense ($n_{pb}/n \sim 0.2$) and slow ($V_{pd} \sim 0.6 V_A$) beam population, with a very anisotropic core, $R_p \gtrsim 3$. Proton beam fraction gradually decreases through S1. The alphas are almost isotropic with a close to negligible beam component, although less isotropic than the times preceding the event, as can be seen by comparing the alpha VDFs in the top two rows in figure \ref{fig:example_vdfs}, and in the slight increase in $\tilde{q}_\alpha$ in figure \ref{fig:event_timeseries}. 
    \item \textbf{S2}: 2020-01-26/14:00:45 to 2020-01-26/14:02:00, a brief, measurable gap between the strong LH and RH polarisation periods, containing 4 SPAN-I fits. The blue LH signature disappears, however upon zooming in there are small flecks of RH polarisation. This is the interval where abrupt changes in beam density fractions and drift speeds in both the protons and the alphas start to occur. 
    \item \textbf{S3}: 2020-01-26/14:02:00 to 2020-01-26/14:16:10, first period of strong coherent (relatively broad band) RH polarisation. Example alpha and proton VDFs from S3 are shown in the third row of figure \ref{fig:example_vdfs}. The proton beam density fraction $N_{pb}/N$ falls to between 0.05 and 0.1, while the drift speed increases to between 1.1 and 1.4$V_A$, constituting a very fast, low density beam. The proton core anisotropy also very suddenly drops from $R_p \sim 3$ to just below $R = 2$. At the same time, a prominent alpha beam abruptly appears, with a density fraction of $n_{\alpha b}/n \sim 0.15$, and a drift speed $V_{\alpha d} \sim 1.2 V_A$. The alpha core anisotropy steps upwards from $R_\alpha \lesssim 1$ to $R_\alpha \sim 1.5$ and continues to increase throughout S3. The change in $R_\alpha$ throughout S3 is much more gradual compared to the step-like changes in $n_{\alpha b}/n$ and $V_{\alpha d}/V_A$. 
    \item \textbf{S4}: 2020-01-26/14:16:10 to 2020-01-26/14:46:25. Remaining period of RH polarisation. From S3 to S4 there is little change in the proton beam and core parameters. The proton beam density fraction remains around $n_{pb}/n \sim 0.05$, with a remarkably steady drift speed around 1.3$V_A$ and the core anisotropy around $R_p\sim 2$. The very fast alpha beam characterising S3 is no longer present in S4, but is abruptly replaced by a slower moving ($V_{\alpha d} \sim 0.4 V_A$) and more dense ($n_{\alpha b}/n \sim 0.5)$ beam. The alpha heat flux $\tilde{q}_\alpha$ remains significantly higher than in S1. No such abrupt changes between S3 and S4 are observed in the core anisotropy $R_\alpha$, it remains elevated throughout this entire period of RH polarisation. Example VDFs from S4 are shown in the fourth row of figure \ref{fig:example_vdfs}.
\end{enumerate}

\section{Fluid Parameters} \label{sec:fluid}
\begin{figure}
    \centering
    \includegraphics[trim= 0cm 0.5cm 0cm 0cm,width=0.9\textwidth]{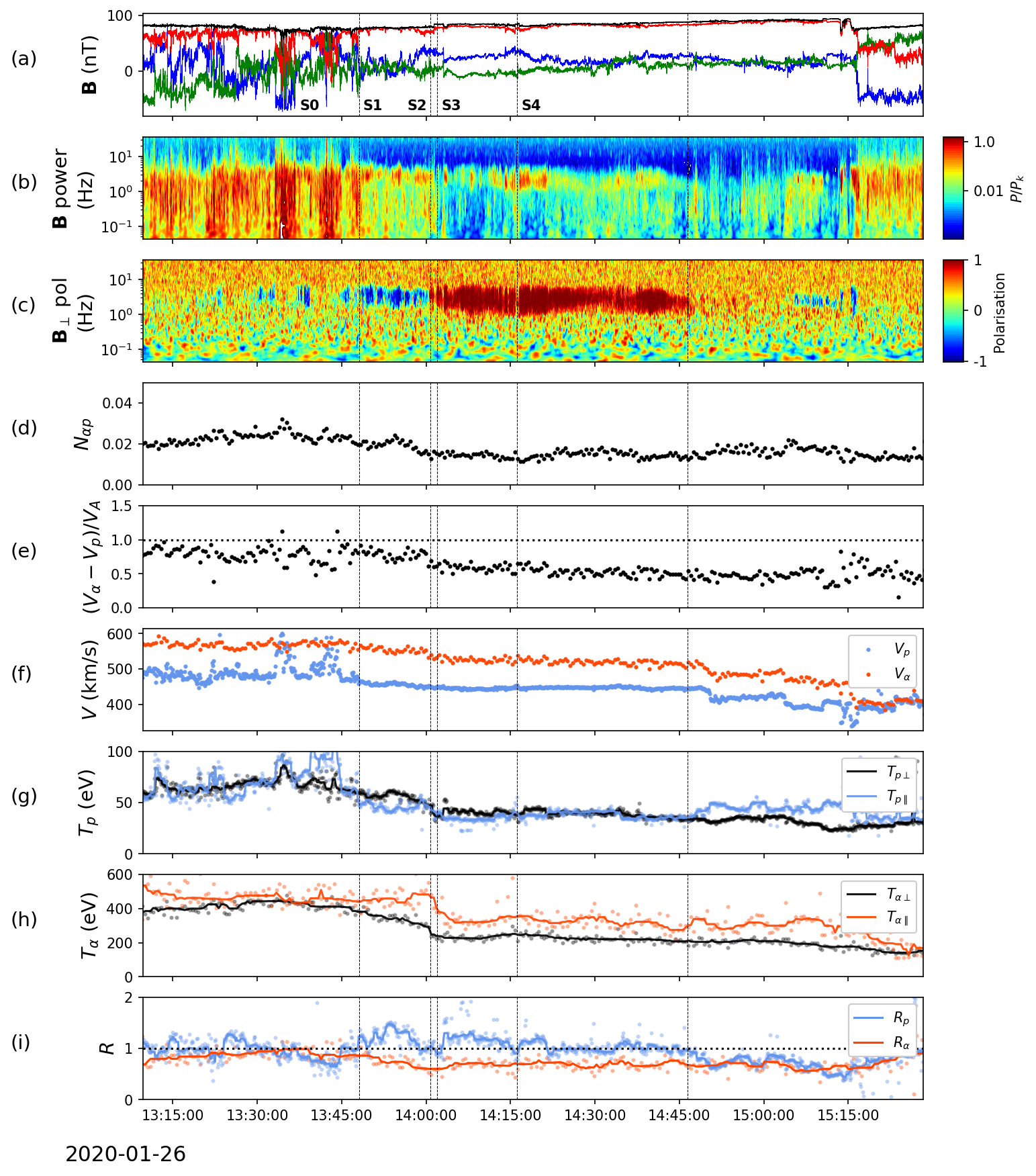}
    \caption{Fluid parameters during the event interval. Panels are (a) magnetic field $\mathbf{B}$ in spacecraft coordinates, (b) magnetic field trace power spectrum, (c) magnetic field perpendicular polarisation, (d) alpha to proton number density ratio, (e) alpha to proton drift speed as a fraction of the Alfvén speed, (f) proton (blue) and alpha (red) centre of mass speeds, (g) proton parallel (blue) and perpendicular (black) temperatures, (h) alpha parallel (red) and perpendicular (black) temperatures, and (i) total temperature anisotropies.}
    \label{fig:fluid_timeseries}
\end{figure}
Figure \ref{fig:fluid_timeseries} shows macroscopic fluid parameters during the interval, in particular the alpha to proton number density ratio $N_{\alpha p}$, alpha-proton drift speed $(V_\alpha - V_p)/V_A$ (where we use the centre of mass velocities for each species $s$, $\mathbf{V}_s = (n_{sc} \mathbf{V}_{sc} + n_{sb}\mathbf{V}_{sb})/(n_{sc} + n_{sb})$, and $V_s = |\mathbf{V}_s|$), the total parallel and perpendicular temperatures $T_\perp,T_\parallel$, and the \emph{total} temperature anisotropies $R_s = T_{s,\perp}/T_{s,\parallel}$. Total temperatures for each species are calculated by combining core and beam temperatures as 
\begin{equation}
    T_\perp = \frac{n_c T_{\perp,c} + n_b T_{\perp,b}}{n_c + n_b}
\end{equation}
\begin{equation}
    T_\parallel = \frac{1}{n_c + n_b} \left( n_c T_{\parallel,c} + n_b T_{\parallel,b} + \frac{n_c n_b}{n_c + n_b}m V_d^2 \right),
\end{equation}
where $V_d$ is the drift speed between the core and beam populations. Neither the alpha abundance nor the alpha-proton drift speed (panels (d) and (e)) change significantly during the wave event itself, with a slight decrease in the alpha-proton drift speed at the start of the RH interval S3, and $N_{\alpha p}$ remaining fairly constant at just below $2\%$. The centre of mass speeds $V_p$ and $V_\alpha$ in panel (f) are remarkably constant, particularly $V_p$. The temperatures of both species (panels (g) and (h)) drop near the beginning of S3, particularly in the alphas, before becoming very steady, while the temperature anisotropies don't vary much between S1 and S4, with $R_\alpha < 1$ and $R_p \gtrsim 1$ throughout. We note that $R_p$ is elevated during S1-S4 relative to the periods immediately before and afterwards, which show no polarisation signature. 

The lack of large variations in the macroscopic drift velocities and temperature anisotropies during the wave event (in contrast to the clear changes in VDF substructure as evidenced in figure \ref{fig:event_timeseries}) strongly suggests that long wavelength fluid instabilities are unlikely to be playing any role, and this is supported by computation of the parameters
\begin{equation}
    \Lambda_M = \sum_s \beta_{s\perp} \left( \frac{T_{s\perp}}{T_{s\parallel}} - 1\right) - \frac{\left(
               \sum_s \rho_s \frac{T_{s\perp}}{T_{s\parallel}}\right)^2}
               {2\sum_s \frac{\rho_s^2}{\beta_{s\parallel}}}
\end{equation}
and 
\begin{equation}
    \Lambda_F = \frac{\beta_\parallel - \beta_\perp}{2} + \frac{\sum_s n_s m_s |\Delta V_s|^2}
         {\sum_s n_s m_s V_A^2},
\end{equation}
associated with the long wavelength mirror \citep{hellinger2007comment,chen2016multi} and firehose \citep{kunz2015inertial} instabilities respectively. Here $\rho_s = q_s n_s$ is the charge density, $\Delta V_s$ the speed relative to the centre of mass, and total $\beta$'s are given by $\beta_{\perp} = \sum_s \beta_{s,\perp}, \beta_{\parallel} = \sum_s \beta_{s,\parallel}$. Given that neither $\Lambda_M$ nor $\Lambda_F$ come close to attaining the instability threshold value of 1 (not shown), we infer that the instabilities active during this period in driving these wave modes unstable are kinetic, not fluid, and moreover, any change in wave observables (such as polarisation) must therefore be associated with the rearrangement of thermal energy between individual core and beam populations in the protons and alphas, rather than any abrupt changes in these large scale plasma parameters. 

\section{Instability Analysis} \label{sec:instability}
To perform the instability analysis we employ the PLUME dispersion relation solver \citep{klein2015predicted,klein2017applying}, which solves the warm plasma dispersion relation for an arbitrary number of drifting bi-Maxwellian distribution functions. In our case we model the plasma with five separate populations: proton core, proton beam, alpha core, alpha beam, and the electrons. Electron density and velocity are calculated from the ion fits in order to enforce charge and current neutrality. The proton core is assigned to be the reference population and all calculations are carried out in this frame. Most importantly, all derived wave frequencies $\omega_r$ will be in the proton core frame. Drift speeds of the other particle populations are defined relative to the reference population and assumed to all lie along $\mathbf{B}$. The set of plasma parameters $\mathcal{P}$ used as input in solving the dispersion relation $\det \mathcal{D}(\omega,\mathbf{k};\mathcal{P}) = 0$ consists of the three parameters defining the reference proton core population:
\begin{equation}
    \mathcal{P}_c = \Biggl\{ \beta_{\parallel,c}, \frac{w_{\parallel,c}}{c}, \frac{T_{\perp,c}}{T_{\parallel,c}}   \Biggr\},\label{eq:core_params}
\end{equation}
where $\beta_{\parallel,c}$ is the parallel plasma beta for the proton core, and $w_{\parallel,c} = \sqrt{2T_{\parallel,c}/m_p}$ the parallel thermal velocity, and a set of 6 dimensionless parameters for each of the proton beam, alpha core, and alpha beam populations, as well as the background electrons:
\begin{equation}
    \mathcal{P}_j = \Biggl\{ \frac{T_{\parallel, c}}{T_{\parallel, j}}, \frac{m_j}{m_p}, \frac{q_j}{q_p}, \frac{T_{\perp,j}}{T_{\parallel,j}},\frac{n_j}{n_c},\frac{dV_j}{V_{Ac}}\Biggr\}.\label{eq:species_params}
\end{equation}
Here $m_j$, $q_j$ and $n_j$ represent the mass, charge, and number density of species $j$ respectively, and $dV_j/V_{ac}$ the drift speed of population $j$ relative to the core, as a fraction of the \textit{core} Alfvén speed:
\begin{equation}
    V_{Ac} = \frac{B}{\sqrt{4\pi m_p n_c}}.
\end{equation}
PLUME employs a signed frequency convention such that $\omega_r > 0$ is forward propagation along $\mathbf{B}$ and $\omega_r < 0$ backward propagation, so that $k_\parallel$ is always positive. We note that although the polarity of the magnetic field during this interval is radially inward (figure \ref{fig:event_timeseries}), as a matter of convention we rectify the field so that positive frequency waves propagate outwards, i.e. anti-sunwards. In terms of PLUME parameters this simply flips the sign of the drift velocities $dV_j$, and makes no physical difference to our interpretation. 

After initially solving the dispersion relation on a grid of $[k_\parallel,k_\perp]$ values, it was found that $\theta^{\text{max}}_{kb}$, the angle between $\mathbf{k}$ and $\mathbf{B}$ for the most unstable mode, was consistently 0 throughout the event. Because of this, we restrict our attention throughout to parallel propagating modes $\mathbf{k} \times \mathbf{B} = 0$ only. Minimum variance analysis of the magnetic field during the event also finds the modes to be parallel propagating, and this is all in concordance with previous statistical studies showing that parallel modes almost always have the highest growth rates \citep{klein2018majority,martinovic2021ion}. We note that under some solar wind conditions, namely a drifting, relatively high density secondary beam population and a lower plasma $\beta_{\parallel,c}$, it is an oblique Alfvén mode that has the lowest instability threshold \citep{daughton1998electromagnetic,gary2000alpha}. For this particular event however, it appears that the beams are not drifting fast enough, nor are high density enough, to be in the region of parameter space where the oblique mode becomes more unstable than the parallel (see for instance figure 9 in \cite{daughton1998electromagnetic} and compare with figure \ref{fig:event_timeseries}). We also note that while we use background electrons here to enforce charge and current neutrality, it is possible with high time resolution data to study the role the electron VDFs play in various wave-particle interactions (for instance \cite{he2022observations} that directly measures the fluctuating electron VDF in response to ion-driven whistler waves in the Earth's foreshock, and \cite{he2020spectra} that investigates the role electrons play in dissipating kinetic Alfvén wave turbulence).



Solving the warm plasma dispersion relation gives us the real frequency and growth rates $\omega = \omega_r + i\gamma$ associated with each normal mode, with the sign $\gamma < 0$ ($\gamma > 0$) denoting plasma stability (instability). The overall power $P$ emitted/absorbed by the plasma can alternatively be computed using the anti-Hermitian part of the dielectric tensor $\epsilon_a$: 
\begin{equation}
    P = \frac{\omega_r}{8\pi} \mathbf{E}^{*} \cdot \epsilon_a \cdot \mathbf{E},
\end{equation}
where $\mathbf{E}$ is the electric field eigenvector for a given mode. Using the linearity of the dielectric tensor and the relation between the dielectric and susceptibility tensors $\epsilon = \mathbb{I} + \sum_s \chi_s$, this is usually written as
\begin{equation}\label{eq:power_abs}
    P \equiv \sum_s \gamma_s = \frac{\omega_r}{8\pi} \sum_s \mathbf{E}^{*} \cdot \chi_{a,s} \cdot \mathbf{E},
\end{equation}
where $\gamma_s$ is therefore the contribution to the overall growth rate associated with species $s$, and $\chi_{a,s}$ is the anti-Hermitian part of the susceptibility tensor for species $s$. This gives a way to discuss the power emitted/absorbed by individual species $s$, and we make use of this extensively in the discussion that follows. 

\subsection{Results}
\subsubsection{S1: LH Polarisation}
\begin{figure}
    \centering
    \includegraphics[trim={0 1cm 0 1cm},width=0.9\textwidth]{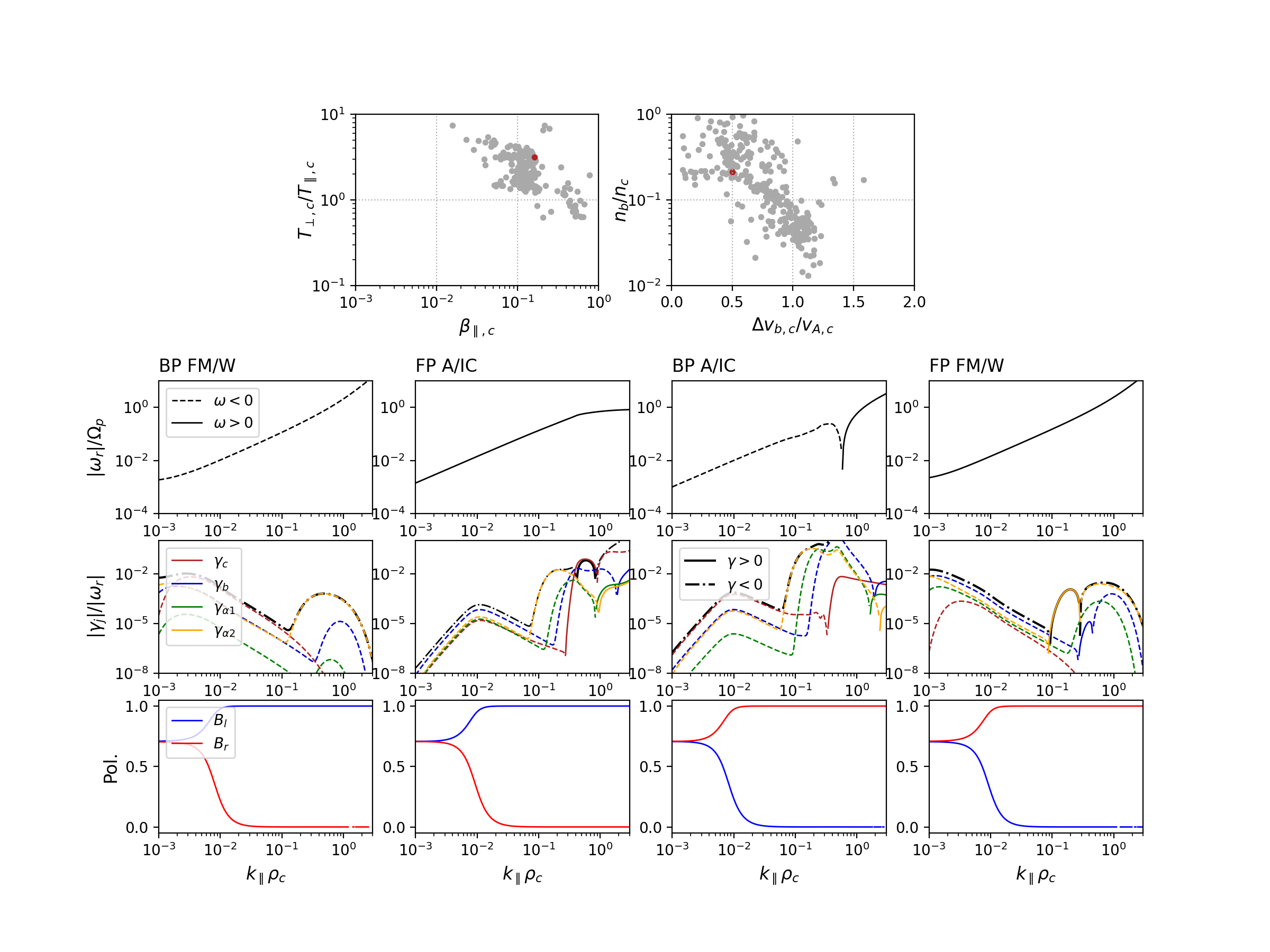}
    \caption{PLUME results for a time slice during interval S1, the period of LH polarisation. Top row shows the distribution of proton core temperature anisotropy $T_{\perp,c}/T_{\parallel,c}$ vs $\beta_{\parallel,c}$, and proton beam density ratio $n_b/n_c$ vs beam drift speed $\Delta v_{b,c}/v_{A,c}$, for the full interval (grey points) and the time slice in question (red point). Bottom three rows show the real frequency $|\omega_r|/\Omega_p$, individual ion component growth rates $|\gamma_j|/|\omega_r|$, and magnetic field polarisation (magnitudes of left and right handed components $B_l$ and $B_r$), as functions of $k_\parallel \rho_c$, for the forward and backward propagating (FP and BP) FM/W and A/IC modes.}
    \label{fig:S1_plume}
\end{figure}
S1 is characterised by the strong, continuous band of LH polarisation from 1 to 5 Hz, and consists of 27 SPAN-fits, meaning we have the results of 27 PLUME calculations. Figure \ref{fig:S1_plume} shows a subset of the PLUME output for a representative time slice in S1 (specifically at 2020-01-26/13:54:23). As we are only focussing on parallel propagating modes, $k_\perp \rho_c$ is fixed to the small value $10^{-3}$, $\rho_c$ the proton gyroradius, and the dispersion relation $\mathcal{D}(\mathbf{k},\omega;\mathcal{P}) = 0$ is solved on a grid of 257 $k_\parallel \rho_c$ values in the range $[10^{-3},3]$ to build up $\omega_r$ and $\gamma$ as functions of $k_\parallel\rho_c$. The results of this are shown in the four columns in the bottom half of figure \ref{fig:S1_plume}. The top row plots the dispersion relation $\omega_r/\Omega_p$ vs $k_\parallel \rho_c$, from which we make the identification of the modes as forward propagating and backward propagating FM/W and A/IC modes. The second row shows the individual contributions to the growth rate from each ion species $\gamma_j$, with the overall growth rate $\gamma$ shown in black. 

For this time slice (typical of all the PLUME results during S1) the forward propagating A/IC mode is seen to be linearly unstable, and is primarily driven unstable by the proton core. That is, the largest contribution to the positive growth rate comes from $\gamma_c$, and is confined to the range $0.3 \lesssim k_\parallel\rho_c \lesssim 0.6$. The effect of the proton beam, alpha core, and alpha beam populations over this $k_\parallel \rho_c$ range is to slightly damp this mode - in figure \ref{fig:S1_plume} the black line denoting overall growth rate $\gamma$ is slightly beneath the $\gamma_c$ curve, and this effect is typical during S1. From the plot of $T_{\perp,c}/T_{\parallel,c}$ vs $\beta_{\parallel,c}$ in figure \ref{fig:S1_plume}, and the 6th panel in figure \ref{fig:event_timeseries}, we conclude that this mode is most likely being driven unstable by a proton-cyclotron instability in the proton core population \citep{verscharen2019multi}. The forward propagating FM/W mode is also predicted linearly unstable at this time, driven by the alpha beam (the orange $\gamma_{\alpha 2}$ curve), albeit with a significantly lower growth rate than that of the A/IC mode. Again, this is typical for S1, with occasional positive contributions to the growth rate from the proton beam population as well. The median dimensionless maximum growth rate of the forward propagating A/IC mode for the entirety of S1 is $\gamma_{\text{max}}/|\omega_r| = 5.1\times 10^{-2}$. The presence of multiple drifting beam populations breaks the symmetry of the VDFs along the magnetic field and so there is no expectation that the forward and backward propagating modes behave similarly \citep{klein2021inferred,podesta2011effect}. Indeed in figure \ref{fig:S1_plume} the two backward propagating A/IC and FM/W modes are both predicted linearly stable at this time and throughout S1 (and in fact for the entirety of the wave event), so we ignore them in the remaining discussion and drop the ``forward propagating" descriptor.


The bottom row of figure \ref{fig:S1_plume} shows the predicted proton core frame wave polarisation by plotting the LH ($B_l$) and RH ($B_r$) magnetic field eigenvector components for each mode as a function of $k_\parallel \rho_c$. In order to compare these polarisations with the magnetic field observations it is necessary to Doppler shift them into the spacecraft frame. We do this explicitly in section \ref{sec:doppler}, but the results of the calculation there show that the Doppler shift does \emph{not} change the sign of $\omega_r$ for the A/IC and FM/W modes and that the polarisation in the proton core frame therefore remains the same in the spacecraft frame. Thus, we expect to observe a blue LH polarisation signature during S1, due to an A/IC mode driven unstable by a proton core temperature anisotropy, which is exactly in agreement with the observations in figure \ref{fig:event_timeseries}.

\subsubsection{S2: Gap Between LH and RH}
\begin{figure}
    \centering
    \includegraphics[width=0.6\textwidth]{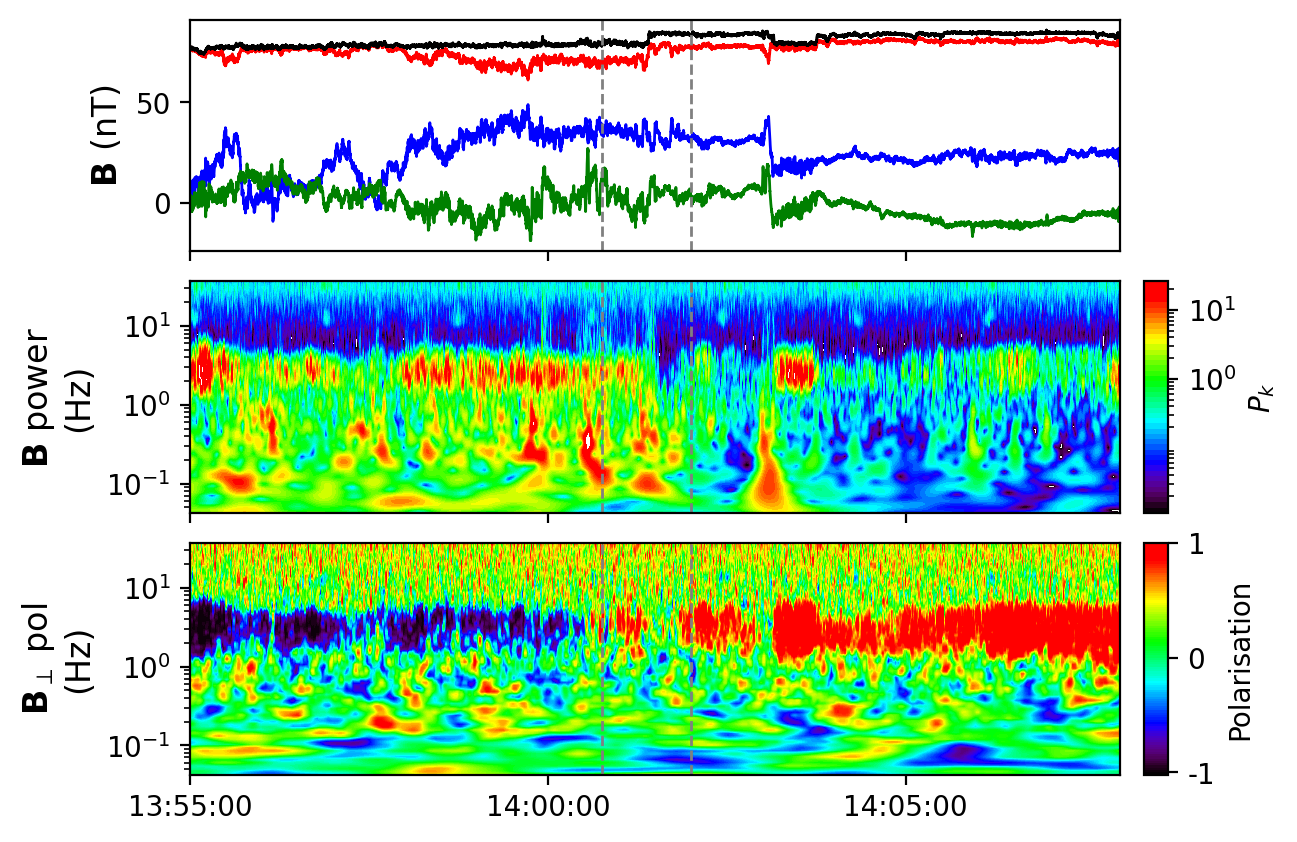}
    \caption{Zoomed in view of the S2 interval (dashed vertical lines), showing the gap between LH and RH polarisation. Top panel: magnetic field in spacecraft coordinates. Middle panel: Trace magnetic field power spectrum. Bottom panel: Magnetic field perpendicular polarisation.}
    \label{fig:S2_zoomed}
\end{figure}
\begin{figure}[!ht]
    \centering
    \includegraphics[width=0.8\textwidth]{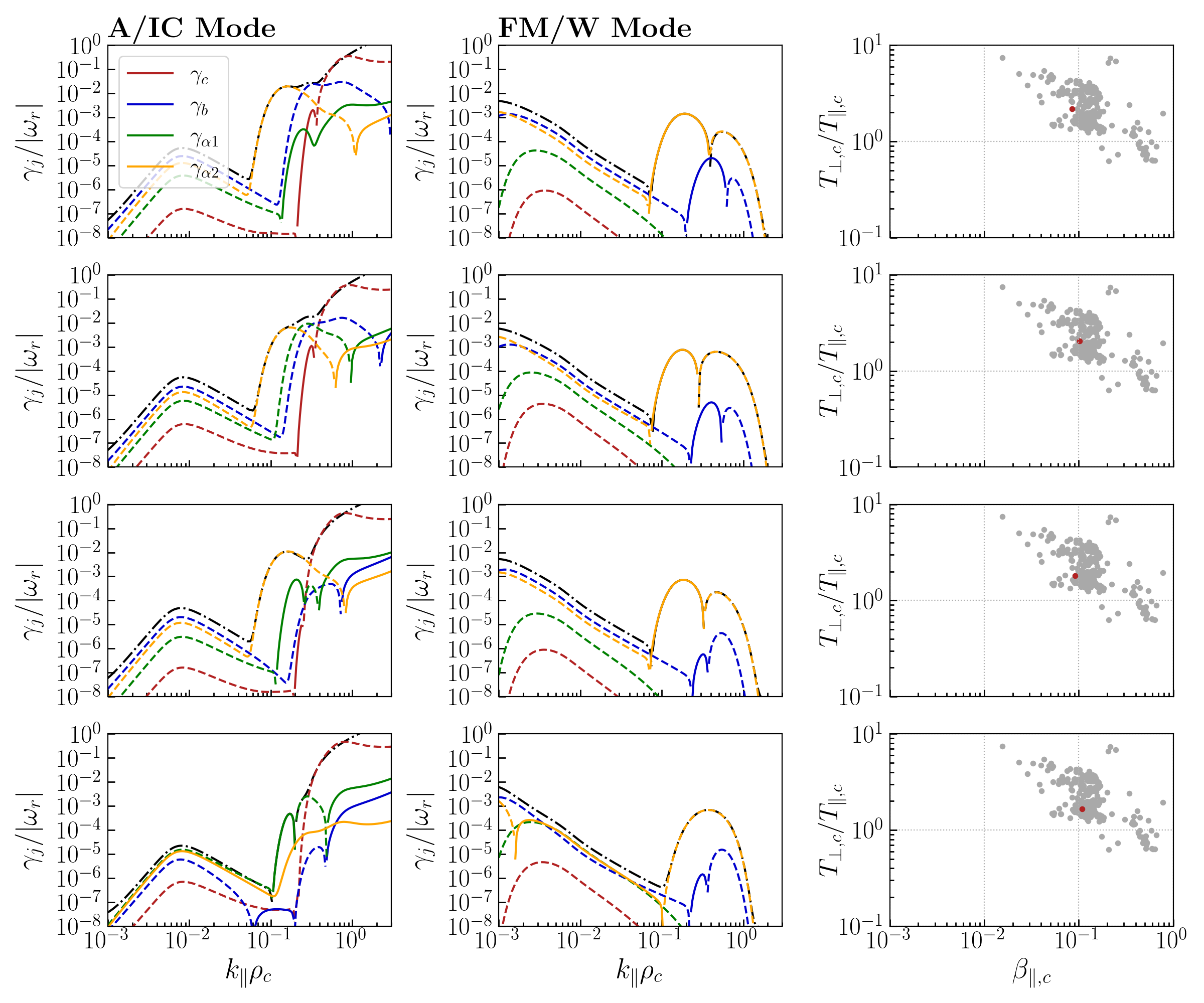}
    \caption{Contributions to the growth rates of the A/IC mode (left column) and FM/W mode (middle column) from each ion species (black dash-dotted lines are the overall growth rate). Each row represents a single time slice during S2. Proton core temperature anisotropies are shown in the right column, with data points from the full interval in grey, and red the measurement at each specific time. The A/IC mode is observed to be linearly stable, coinciding with the dropout in LH polarisation seen in figure \ref{fig:S2_zoomed}.}
    \label{fig:S2_growth_rates}
\end{figure}
A zoomed in view of interval S2, the short gap between the bands of LH and RH polarisation, is shown in figure \ref{fig:S2_zoomed}. From this, we can see that the LH signature in panel 3 has disappeared completely, but there are still small measurable flecks of RH polarisation present. There is also a fairly marked drop off in power in the frequency range of interest (panel 2). In S2 we have 4 SPAN-I fits and so 4 PLUME results to analyse. Each row of figure \ref{fig:S2_growth_rates} shows the ion component growth rates for both the A/IC mode and the FM/W mode for each of these 4 time slices. Remarkably, we can see that the A/IC mode (first column) is no longer unstable, coinciding exactly with the disappearance of LH polarisation. Comparing figure \ref{fig:S2_growth_rates} to figure \ref{fig:S1_plume}, the damping rates from the components other than the proton core are all broadly similar, and it is a drastic reduction in the proton core growth rate $\gamma_c$ that is responsible for the mode becoming overall linearly stable. The right hand column shows the proton core temperature anisotropy as a function of parallel core $\beta_{\parallel,c}$, and comparing this to its equivalent plot in figure \ref{fig:S1_plume} (and the data in figure \ref{fig:event_timeseries}) it is clear there is a reduction in the proton core anisotropy during S2, further supporting our interpretation in the previous section that this mode is being driven unstable by a proton cyclotron instability. The middle column in figure \ref{fig:S2_growth_rates} shows the growth rates for the FM/W mode, and similarly to the time slice in figure \ref{fig:S1_plume}, it is unstable, predominantly driven so by the alpha beam population. 

\subsubsection{S3: Initial Period of RH Polarisation}
\begin{figure}
    \centering
    \includegraphics[trim={0.5cm 1.5cm 0.5cm 0},width=0.75\textwidth]{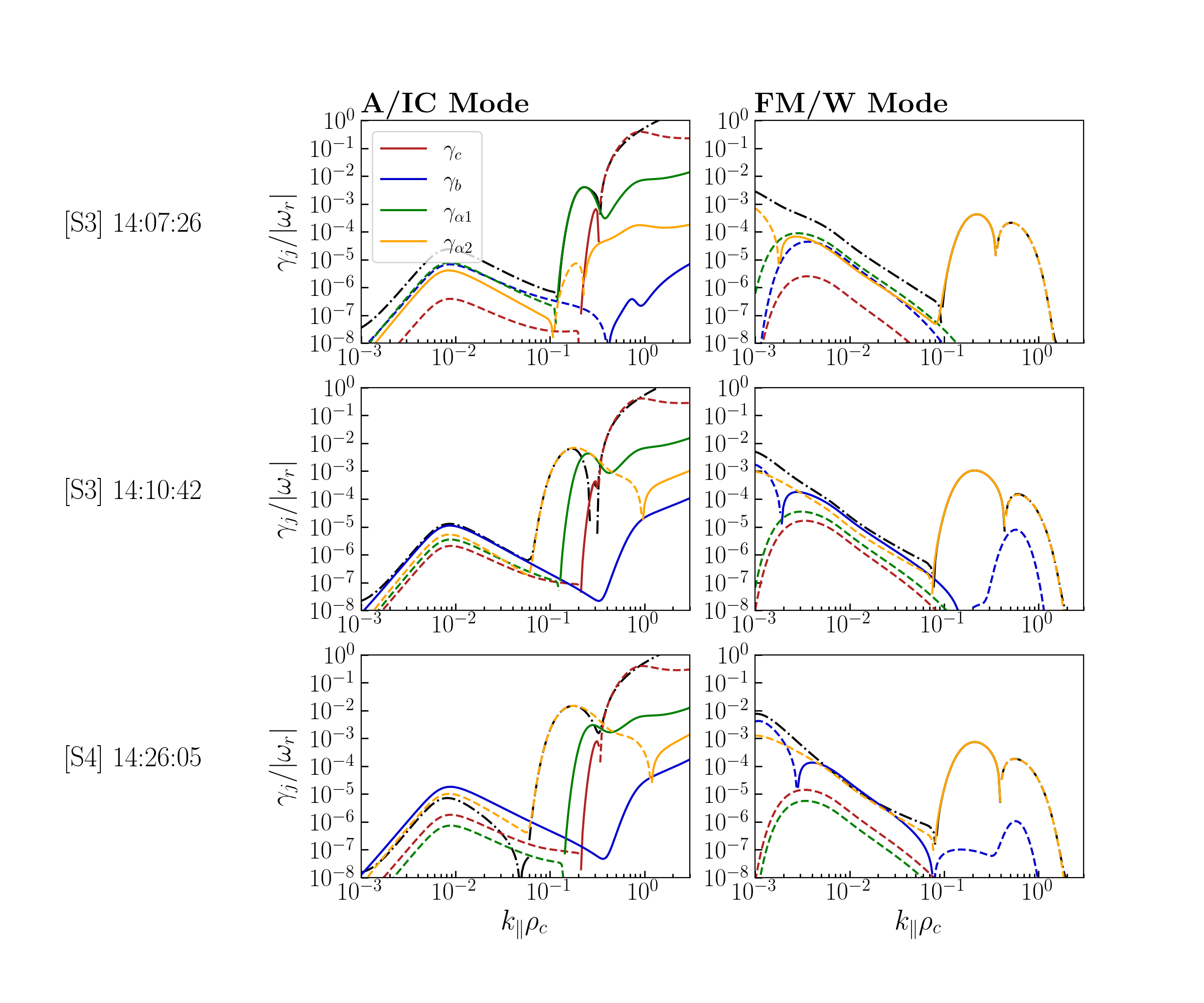}
    \caption{Example growth rates of the A/IC and FM/W modes during two different times in subinterval S3 (top two rows) and one time during S4 (bottom row), illustrating how the A/IC mode in S3 is sometimes predicted to be driven unstable by the alpha core population (green), while at other times, and during S4, it is heavily damped by the alpha beam (orange). The FM/W mode on the other hand is consistently driven unstable by the alpha beam throughout.}
    \label{fig:S3_S4_growthrates}
\end{figure}
In contrast to S1 and S2 (and, as we shall see, S4), S3 is less straightforward to understand, and the agreements between the observations and the PLUME predictions are not as good. The PLUME results for maximum growth rate are of two types, shown in the top two rows of figure \ref{fig:S3_S4_growthrates}. In the first, the A/IC mode is predicted to be linearly unstable with a higher growth rate than the FM/W mode, with the dominant positive contribution to $\gamma$ coming from the alpha core component $\gamma_{\alpha 1}$ (green), in contrast to the proton core component during S1. In the second, the FM/W mode is unstable and the A/IC mode stable, and is again being driven by the alpha beam component $\gamma_{\alpha 2}$. Of the 30 PLUME results in S3, 16 of them have $\gamma^{A/IC} > \gamma^{FM/W}$, and 13 of them have $\gamma^{FM/W} > \gamma^{A/IC}$. We would therefore expect to see a roughly equal mixture of LH and RH polarisations, which is clearly at odds with the purely red RH band we observe throughout. The median maximum growth rate of these A/IC modes is $\gamma^{A/IC}/|\omega_r| = 4.2 \times 10^{-3}$, lower than the corresponding growth rate in S1, and that of the FM/W modes is $\gamma^{FM/W}/|\omega_r| = 5.2 \times 10^{-4}$. 

\subsubsection{S4: Second Period of RH Polarisation}
S4 is characterised by the abrupt disappearance of the very fast moving, tenuous alpha beam seen during S3, replaced by a slower moving, denser one ($V_{\alpha d} \sim 0.4V_A, n_{\alpha b}/n \sim 0.5$). A representative plot of the A/IC and FM/W growth rates and their individual species contributions during S4 is shown in the third row of figure \ref{fig:S3_S4_growthrates}. As in the previous three sub-intervals the FM/W mode is being driven unstable by the alpha beam population. Of the 64 PLUME outputs in S4, 38 of them have $\gamma^{FM/W} > \gamma^{A/IC}$, and only 9 of them have $\gamma^{A/IC} > \gamma^{FM/W}$ (here $\gamma$ implicitly means $\gamma_{\text{max}})$. Thus, we would expect S4 to be majority RH polarised, which is in agreement with the observations. The median growth rate for the FM/W mode is $\gamma/|\omega_r| = 3.5 \times 10^{-4}$, and the median growth rate of the A/IC mode is $\gamma/|\omega_r| = 4.0 \times 10^{-3}$. We also point out that for the majority of times during S4 where the A/IC max growth rate is larger than the FM/W max growth rate, the range of $k_\parallel \rho_c$ over which the mode is unstable is exceedingly narrow, suggesting a degree of somewhat delicate cancellations between the positive and negative growth rate contributions.  

We note that, although there is some tension between the observations and PLUME predictions during S3, in all cases it is the alphas - core and beam - that are playing the primary role in determining plasma stability here, with the proton contributions being subdominant. Alpha associated modes are also unstable at smaller wavevectors $k_\parallel\rho_c$, so regardless of polarisation we predict that the wave frequencies during the RH periods S3 and S4 are lower than those during the LH period S1, which we examine in section \ref{sec:doppler}.

\subsection{Doppler Shift Calculation}\label{sec:doppler}
\subsubsection{Polarisation}
In this section we perform the Doppler shift calculation of the proton core frame frequencies and intrinsic polarisations into the spacecraft frame, in order to compare the PLUME predictions to the magnetic field observations. If $(\omega_r,\mathbf{k})$ are the real frequency and wave vector associated with a given mode in the proton core frame, then the Doppler shifted frequency observed in the spacecraft frame, $\omega_{\text{sc}}$, is given by
\begin{equation}
    \omega_{\text{sc}} = \omega_r + \mathbf{k}\cdot \mathbf{V}_{\text{sw}},\label{eq:doppler_shift}
\end{equation}
where $\mathbf{V}_{\text{sw}}$ is the solar wind velocity (in this case the proton core velocity). Here we use the $\omega_r > 0$ convention, with $\mathbf{k}$ denoting the direction of wave propagation. From equation \ref{eq:doppler_shift}, since $\mathbf{V}_{\text{sw}}$ is always radially outwards, waves that also propagate outwards are always shifted to higher frequencies $\omega_{\text{sc}} > \omega_r$, as $\mathbf{k}\cdot \mathbf{V}_{\text{sw}} > 0$. The polarisation of outward propagating waves therefore remains the same in the spacecraft frame as it was in the proton core frame. For radially inward propagating waves, $\mathbf{k}\cdot\mathbf{V}_\text{sw} < 0$, there are two possibilities. The first is that the shift to lower frequencies $\omega_{\text{sc}} < \omega_r$ leaves $\omega_{\text{sc}} > 0$. In this case the spacecraft frame polarisation again remains unchanged relative to the proton frame polarisation. The second possibility is that the Doppler shift is large enough to make $\omega_{\text{sc}} < 0$. These inwardly propagating waves will be measured to have the opposite polarisation to their proton frame polarisation. In addition, since the spacecraft measured frequency is a positive definite quantity, these waves will have a measured frequency of $|\omega_\text{sc}| = -\omega_\text{sc}$ in the spacecraft frame (see \cite{bowen2020electromagnetic} for an in-depth discussion of these ambiguities and their relation to in-situ observations).

As discussed above, the relevant modes as predicted by PLUME here are the outward propagating A/IC and FM/W modes, which will therefore be Doppler shifted to higher frequencies in the spacecraft frame, retaining their intrinsic proton frame polarisations with no possibility of a sign change in $\omega_{\text{sc}}$. Thus our interpretation of the wave event as being an A/IC mode during S1, the disappearance of the A/IC mode during S2, and predominantly an FM/W mode during S3 and S4, predicts a change from LH polarisation during S1, to no LH polarisation in S2, to RH polarisation during S3 and S4, which is exactly what is observed. This is summarised in the middle panel of figure \ref{fig:summary_plot}, where the maximum growth rate of the most unstable mode is plotted as a function of time, coloured by its spacecraft frame polarisation. Faded red dots represent times when both the FM/W and A/IC modes are unstable, but $\gamma^{FM/W} < \gamma^{A/IC}$. 


\subsubsection{Predicted Frequencies}
\begin{figure}
    \centering
    \includegraphics[width=0.5\textwidth]{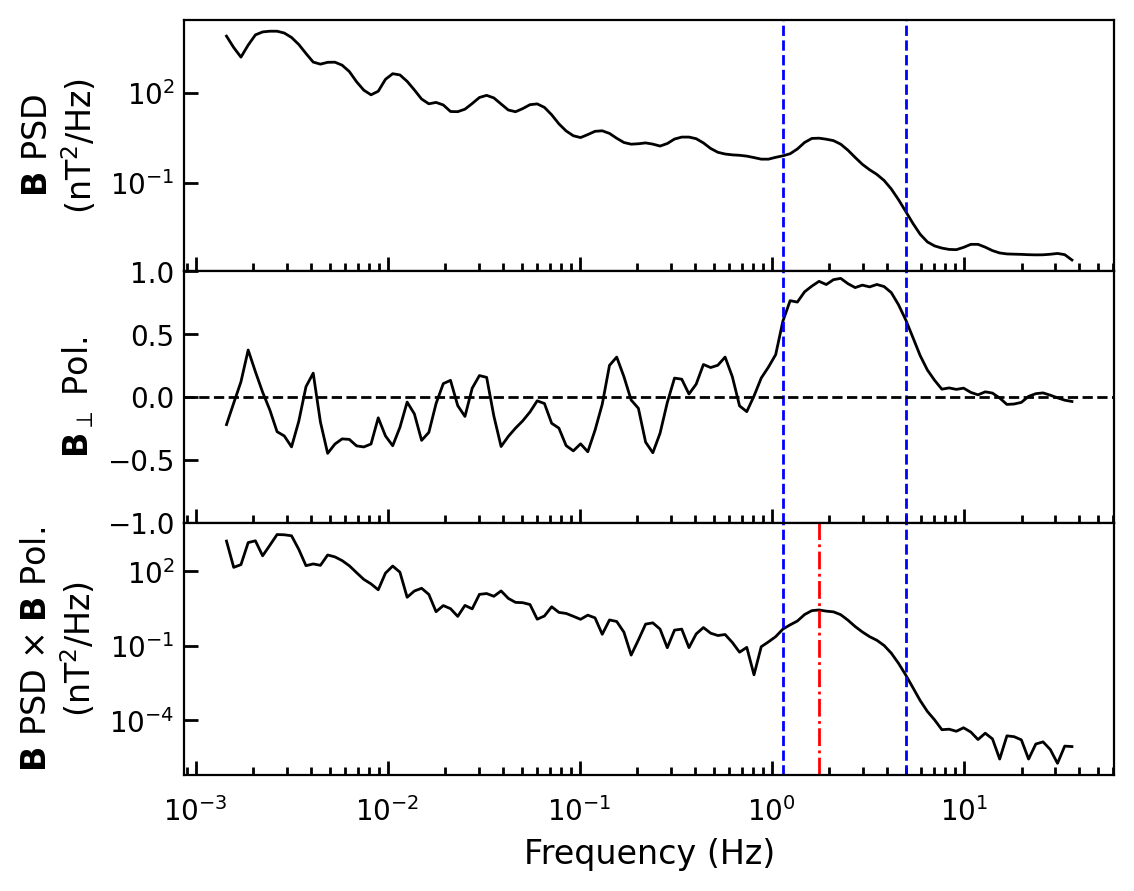}
    \caption{Example illustrating how the frequency of the observed ion scale wave is estimated. Top panel shows magnetic field power spectral density with characteristic bump at 2 Hz indicating coherent wave power, middle panel is perpendicular $\mathbf{B}$ polarisation, bottom panel is the product of the two. Wave frequency is estimated as the maximum of this product (red line) in the range where $|\mathbf{B}_\perp \text{ Pol}| > 0.5$ (blue lines).}
    \label{fig:central_freq_est}
\end{figure}

For a given unstable mode, we can make a prediction of its observed frequency by finding the $(\omega_r,k_\parallel)$ corresponding to its maximum growth rate $\gamma^{\text{max}}$, and substituting these into equation \ref{eq:doppler_shift}. Since we are restricting the calculation to parallel propagation $\mathbf{k}\times\mathbf{B} = 0$, $\mathbf{k}\cdot\mathbf{V}_\text{sw} = k_\parallel \hat{\mathbf{b}} \cdot \mathbf{V}_{\text{sw}}$. 
To estimate the central frequency associated with the observed bands of polarisation in $\mathbf{B}_\perp$, we find the frequency at which the product of the magnetic field power spectrum and the polarisation is at a maximum, restricted to the frequency rage over which the polarisation is larger than 0.5 in magnitude; an example is shown in figure \ref{fig:central_freq_est}. This method therefore makes use of information from both available observational features. 
\begin{figure}
    \centering
    \includegraphics[width=0.7\textwidth]{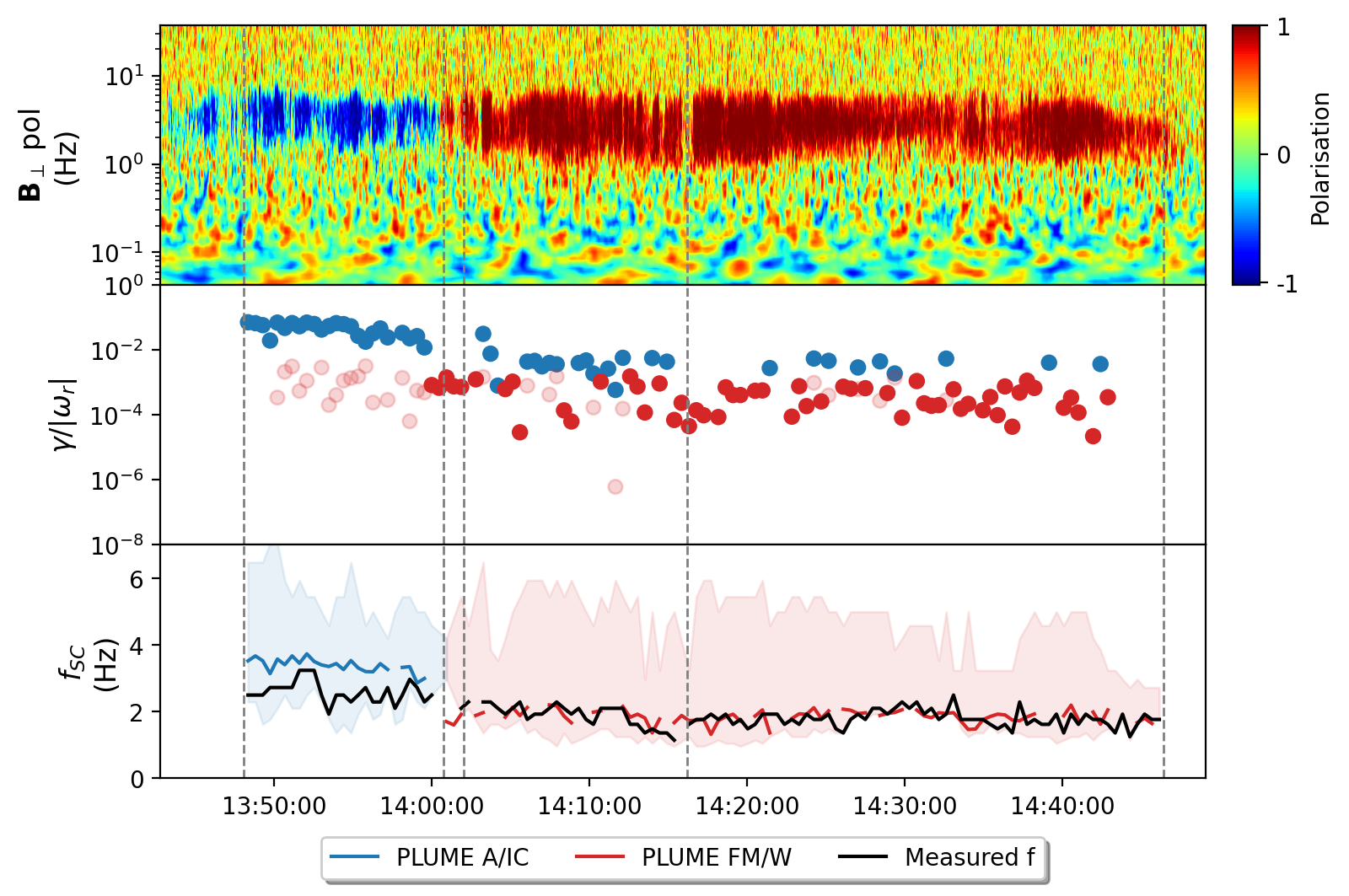}
    \caption{Summary of the wave event showing $\mathbf{B}_\perp$ polarisation (top panel), maximum growth rates coloured by predicted spacecraft frame polarisation (blue for the A/IC mode and red for the FM/W mode) (middle panel), and measured ion scale wave frequencies (black) compared to PLUME predicted frequencies in blue (A/IC mode) and red (FM/W mode) (bottom panel). Faded red dots in the second panel indicate $\gamma_{\text{max}}$ of the FM/W mode that is also unstable but smaller in magnitude than $\gamma_{\text{max}}$ of the A/IC mode. The faded blue and red envelopes in the third panel represent the frequency range over which $|\mathbf{B}_\perp \text{ pol}| > 0.5$.}
    \label{fig:summary_plot}
\end{figure}

The comparison between the PLUME predicted frequencies and the observed frequencies are shown in the bottom panel of figure \ref{fig:summary_plot} as a function of time, where we have taken the A/IC mode as the predominant mode during S1, and the FM/W mode for the remainder of the interval. The agreement between the predicted and observed frequencies is remarkably good, with a difference of $\lesssim 1$ Hz during the LH period, and the two time series lying essentially on top of each other during the RH interval. For those times in S3 and S4 where the A/IC mode is predicted most unstable rather than the FM/W mode, the associated predicted frequency is generally higher than the FM/W mode (up to about 0.8 Hz higher, not shown), which would therefore constitute a worse agreement with the measured wave frequencies in figure \ref{fig:summary_plot}. This, together with the observed polarisation being only RH, reinforces the idea that these times are erroneous predictions and the FM/W mode really is the only relevant one during S3 and S4. We discuss some possible reasons for this discrepancy in section \ref{sec:conclusions}. We note that while the agreement in figure \ref{fig:summary_plot} looks very good, different methods of estimating the central observed frequency do give different results on the order of 1-2 Hz. Although the agreement is less good during the LH interval, the fact that the observed decrease in wave frequency between the LH and RH intervals is clearly reflected in the PLUME predictions further supports our interpretation of the LH and RH waves as being proton and alpha associated modes respectively. 

\subsection{Parameterisations}\label{sec:paramterisations}
\begin{figure}
    \centering
    \includegraphics[width=0.9\textwidth]{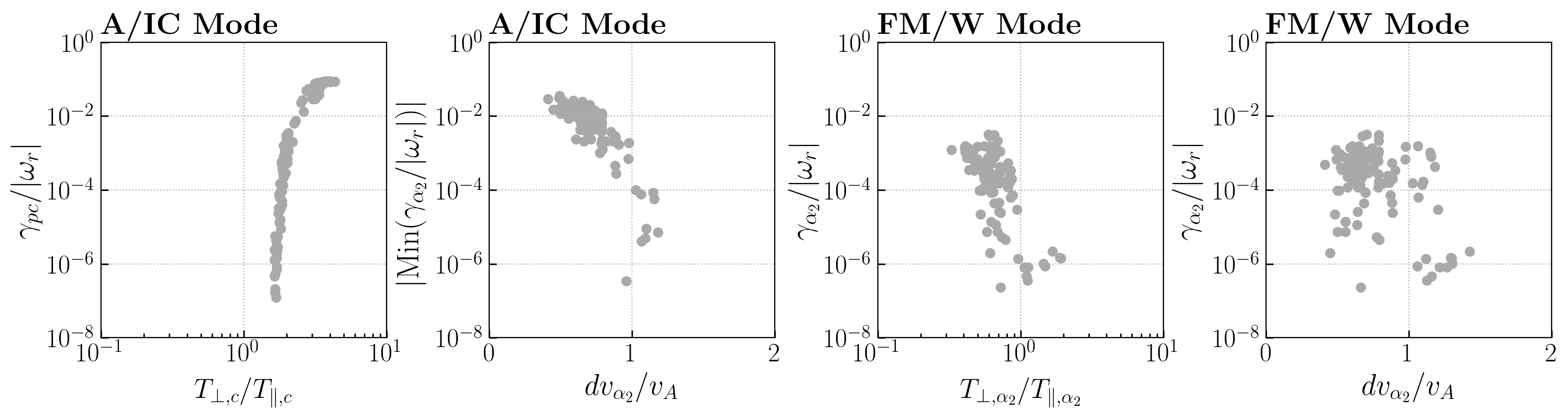}
    \caption{From left to right: Proton core contribution to the normalised growth rate of the A/IC mode vs proton core temperature anisotropy; damping rate of the alpha beam component on the A/IC mode as a function of alpha beam drift speed; alpha beam contribution $\gamma_{\alpha_2}$ to the FM/W mode vs the alpha beam temperature anisotropy; $\gamma_{\alpha_2}$ vs alpha beam drift speed.}
    \label{fig:individual_growth}
\end{figure}
As discussed in the introduction, statistical studies of kinetic instabilities in the solar wind often focus on 2D histograms as functions of $R_p$ and $\beta_\parallel$ (``Brazil plots"). These usually only consider protons, but each new species adds several additional parameters to the system. Indeed, one possible explanation of why the oblique instability thresholds seem to do a better job of constraining the measurements \citep{hellinger2006solar,bale2009magnetic} despite the fact that empirically in case studies the waves usually seem to be parallel propagating could be the sensitivity of these Brazil plots to other effects, such as additional species and their relative drifts \citep{gary2003consequences,hellinger2006parallel,klein2018majority,martinovic2021ion,vafin2018solar} or non-Maxwellian distribution functions \citep{verscharen2018alps,pierrard2010kappa}. 
In reality, these systems are high-dimensional with correlations existing between different plasma parameters, for instance beam drift speeds and beam density ratios (see e.g. figure \ref{fig:S1_plume} or figure 9 in \cite{klein2021inferred}), and overall stability may depend sensitively on many different quantities at once. Simple parameterisations of instability thresholds as a function of only two variables, such as those in Brazil plots, would therefore not a priori be expected to constrain solar wind measurements well. 

In figure \ref{fig:individual_growth} we show several different parameterisations of the A/IC and FM/W mode growth rates (the data points are from S1 through S4). The first plot shows the proton core normalised growth rate $\gamma_{pc}/|\omega_r|$ as a function of the proton core temperature anisotropy, showing a remarkably clear positive correlation. Given that $\gamma_{pc}$ is the species with the highest positive contribution to the growth rate during S1, this confirms our interpretation above that the LH polarisation signatures we see are LH A/IC waves driven by a proton core cyclotron instability. The second plot is the maximum \emph{damping} rate on the A/IC mode due to the alpha beam, showing a clear negative correlation; slower moving alpha beams are responsible for greater damping of the forward propagating A/IC mode. The very fast moving alpha beam during S3 does not sufficiently damp the forward propagating A/IC mode, producing the conflicting prediction of the most dominant wave mode during this period. During S4, when the alpha beam becomes slower and more dense, the A/IC mode is more heavily damped, allowing the FM/W mode to dominate.

The next two plots show the dependence of the alpha beam contribution $\gamma_{\alpha_2}/|\omega_r|$ to the growth rate of the FM/W mode on the temperature anisotropy $T_{\perp,\alpha_2}/T_{\parallel,\alpha_2}$ and drift speed $dv_{\alpha_2}/v_A$. There is a negative correlation with the temperature anisotropy, but the relation with the drift speed is much less obvious (if indeed there is one at all). One may conclude that it is therefore a firehose instability $T_{\perp,\alpha_2}/T_{\parallel,\alpha_2} < 1$ responsible for driving the RH wave mode being measured, as opposed to an alpha-proton drift instability. We point out though that the parallel firehose instability and the ion/ion right handed resonant instability are really two limiting cases of a single generalised instability, that  associated with a drifting beam allowed to be anisotropic \citep{verscharen2013parallel}. A pure parallel firehose instability would have $V_d = 0$ and $T_\perp/T_\parallel < 1$, while a pure ion/ion instability has $T_\perp/T_\parallel = 1$ and $V_d \neq 0$. FM/W waves are driven by an excess of parallel pressure which can be supplied by either a temperature anisotropy or a drift along $\mathbf{B}$.

\section{Conclusions} \label{sec:conclusions}
In this paper we have studied in detail an ion scale wave event from PSP’s E4, identified as a period of enhanced magnetic field fluctuations accompanied by a $\sim 15$ minute period of LH magnetic field perpendicular polarisation followed by a very abrupt change to a strong $\sim 45$ minute interval of RH polarisation. SPAN-I observations during the event showed clear secondary populations in the proton and alpha VDFs, leading us to fit core and beam bi-Maxwellians to both species. These were then input into the PLUME warm plasma dispersion relation solver \citep{klein2015predicted} to identify and characterise the four most linearly unstable (or least damped) modes at each time slice. The LH polarised waves were identified as forward propagating (radially outwards) A/IC waves, primarily driven by a proton cyclotron instability $T_{\perp,c}/T_{\parallel,c} > 1$ in the proton core population, with a median growth rate of $\gamma/|\omega_r| = 5.1 \times 10^{-2}$. We found that the A/IC mode was damped by the proton beam, the alpha core, and the alpha beam populations, with the largest damping generally being supplied by the alpha beam. This damping effect of the alphas on proton excited wave modes was also observed and remarked upon in the 2.5D and 3D expanding box simulations of \cite{ofman2022modeling}, which used SPAN-I proton and alpha parameters from this same E4 event as initial conditions. The RH polarised waves were identified as forward propagating FM/W waves, primarily driven by a firehose-like $T_{\perp,\alpha_2}/T_{\parallel,\alpha_2} < 1$ instability in the alpha beam population, with a significantly lower median growth rate of $\gamma/|\omega_r| = 3.5 \times 10^{-4}$. The abrupt change from LH to RH polarisation was caused by a decrease in the proton core temperature anisotropy, drastically reducing the growth rate of the A/IC mode and allowing the formerly sub-dominant alpha beam driven FM/W mode to dominate. In the very short gap immediately after the LH signature first drops out, we showed the plasma becoming rapidly stable to the A/IC mode. The spacecraft frame Doppler shifted wave frequencies and polarisations agreed extremely well with the magnetic field measurements, with the main discrepancy being the prediction that the first sub-interval of the RH period should be a roughly equal admixture of LH and RH modes. Our fits during this sub-interval showed a very fast and tenuous alpha beam component which was unable to sufficiently damp the forward propagating A/IC mode (due to its large drift speed relative to the proton core), leading to the erroneous prediction of LH polarisation. We note that normalising the growth rates $\gamma$ to the same absolute time scale $\Omega_p^{-1}$, rather than $\omega_r^{-1}$ (which is slightly different between the fast and Alfvén modes), still does not resolve this tension. One possible reason for this discrepancy between the PLUME predictions and the observations is that modelling ion VDFs by two bi-Maxwellians is of course only an approximation to their true shape.  Wave-particle interactions via cyclotron resonances lead to velocity space diffusion in the vicinity of the resonance and the production of ``diffusion plateaus" \citep{tu2002anisotropy,heuer2007diffusion,cranmer2014ensemble,he2015evidence}. Depending on the location of the resonant velocity, if some degree of diffusion has already occurred (as is likely the case, see \cite{bowen2022situ} and \cite{verniero2022strong}) then a bi-Maxwellian fit can significantly over-estimate the pitch-angle gradient (being unable to capture a flat region of the VDF), predicting a higher growth rate $\gamma$ than the true VDF would support. To fully answer whether this is what is happening here would require carefully looking at fit residuals as a function of velocity, or solving the warm plasma dispersion relation for the ``raw" VDF (using software packages such as ALPS \citep{verscharen2018alps} or LEOPARD \citep{astfalk2017leopard}), as opposed to decomposing them into bi-Maxwellians (see \cite{walters2023effects} for a recent example of this approach). We leave these questions as avenues for future investigation. Overall however, the degree of agreement between the observations and the predicted spacecraft frame polarisations, wave frequencies, and times of occurrence and disappearance, strongly suggest we are measuring locally generated ion scale waves caused by the non-Maxwellian features observed in the alpha and proton VDFs. 

Our identification of the ion-scale wave modes active during this event as all being anti-Sunward propagating is in agreement with previous case studies \citep{gary2016ion,verniero2020parker}, and statistical surveys \cite{klein2018majority,bowen2020electromagnetic,martinovic2021ion} which concluded that the majority of observed ion scale waves are propagating outward, thus maintaining their intrinsic plasma frame polarisations. Similarly, previous results \citep{jian2009ion,bowen2020electromagnetic} have suggested that LH events are more frequent than RH ones. Looking at the entirety of this 4th PSP encounter there is a clear pattern of majority LH polarised events during the inbound portion (with this event being a conspicuous exception), however the outbound portion is dominated by RH polarised wave signatures. Determining the reason for this change on either side of perihelion (notably PSP dropped into a slow speed stream on the 30th), and determining what other drivers there are for generating RH polarised waves  at these radial distances (i.e. whether the alphas are usually involved) would be worthwhile avenues of further work. The observation of a secondary alpha population, and the fact that it was playing an active role in the dynamics (being responsible for driving the plasma linearly unstable) is a novel observation given that alpha beams are somewhat rarely remarked upon in the literature. Whether such wave events involving secondary alpha populations are rare or not, and what relationship (if any) they have with their proton beam counterparts are also interesting questions for further study. 


\section{Appendix}
\begin{figure}
    \centering
    \includegraphics[width=0.8\textwidth]{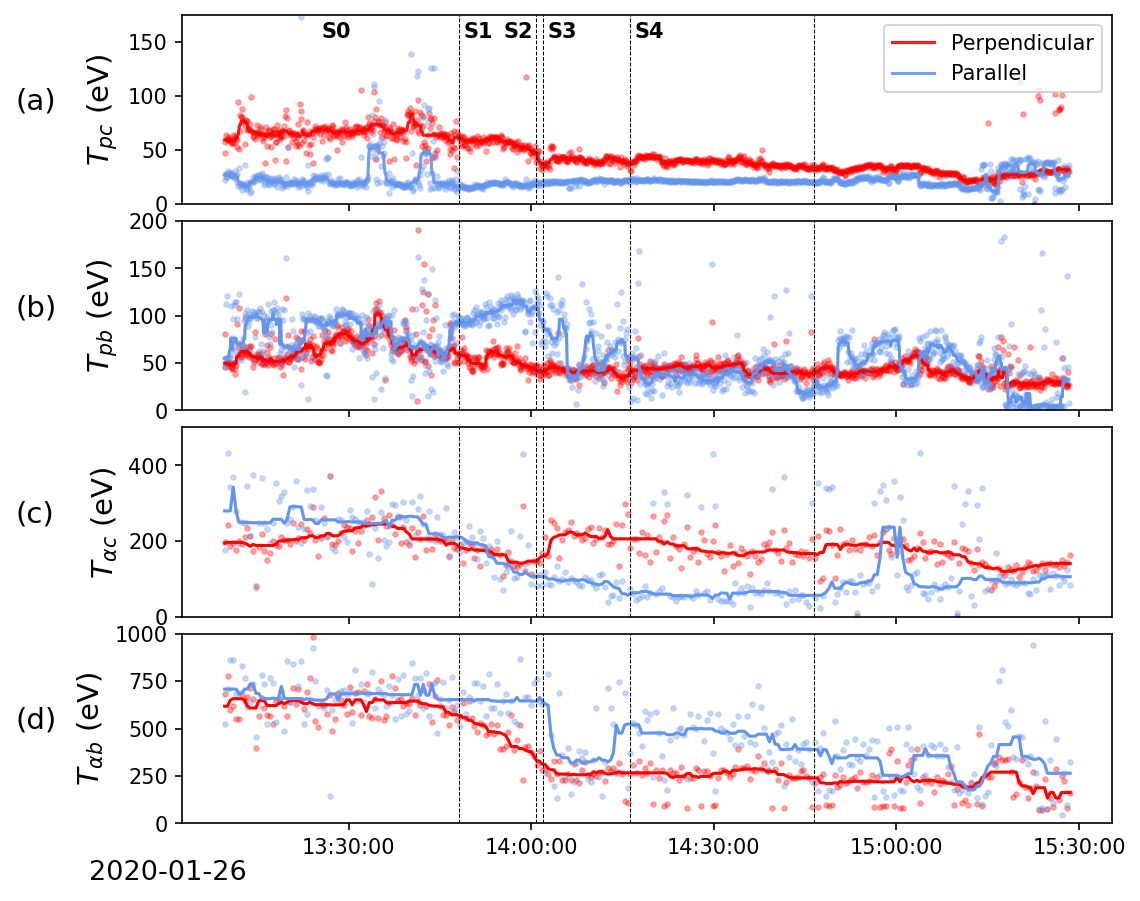}
    \caption{Time series of the perpendicular (red) and parallel (blue) temperatures of the (a) proton core, (b) proton beam, (c) alpha core, (d) alpha beam populations. Vertical dashed lines demarcate the sub-intervals S0 through S4 as defined in the text.}
    \label{fig:temp_timeseries}
\end{figure}
For completeness in figure \ref{fig:temp_timeseries} we plot the individual perpendicular and parallel temperatures of each ion component (proton core, proton beam, alpha core, alpha beam), and in table \ref{tab:params} we list the various plasma population parameters $\mathcal{P}_j$ (c.f. equations \ref{eq:core_params} and \ref{eq:species_params}) for figures \ref{fig:S1_plume}, \ref{fig:S2_growth_rates}, and \ref{fig:S3_S4_growthrates} respectively. For brevity the trivial mass and charge ratios are not included. 

\begin{table}
\caption{Plasma parameters for figures} \label{tab:params}
\begin{tabular}{c c c c c c c c c}

    & $\mathcal{P}_j$  & $\beta_{\parallel,c} $ & $\frac{w_{\parallel,c}}{c}$ & $\frac{T_{\perp,c}}{T_{\parallel,c}}$ & $\frac{T_{\parallel,c}}{T_{\parallel,j}}$ & $\frac{T_{\perp,j}}{T_{\parallel,j}}$ & $\frac{n_j}{n_c}$ & $\frac{dV_j}{V_{Ac}}$\\ 
   
    \hline
    \hline

    \multirow{4}{*}{Fig. 4} & $\mathcal{P}_c$ & 0.162 & 0.000204 & 3.147 & - & - & - & -\\ 
    & $\mathcal{P}_b$ & - & - & - & 0.215 & 0.656 & 0.213 & 0.501\\
    & $\mathcal{P}_{\alpha 1}$ & - & - & - & 0.0977 & 0.898 & 0.0158 & 0.618\\ 
    & $\mathcal{P}_{\alpha 2}$ & - & - & - & 0.0266 & 0.707 & 0.0187 & 0.680\\
    
    \hline

    \multirow{4}{*}{Fig. 6 (a)} & 
    $\mathcal{P}_c$ & 0.0849 & 0.000201 & 2.188 & - & - & - & -\\ 
    & $\mathcal{P}_b$ & - & - & - & 0.167 & 0.342 & 0.175 & 0.402\\
    & $\mathcal{P}_{\alpha 1}$ & - & - & - & 0.234 & 1.818 & 0.0142 & 0.384\\ 
    & $\mathcal{P}_{\alpha 2}$ & - & - & - & 0.0248 & 0.406 & 0.0180 & 0.589\\
    
    \hline
    \multirow{4}{*}{Fig. 6 (b)} & 
    $\mathcal{P}_c$ & 0.102 & 0.000203 & 2.040 & - & - & - & -\\ 
    & $\mathcal{P}_b$ & - & - & - & 0.214 & 0.416 & 0.116 & 0.454\\
    & $\mathcal{P}_{\alpha 1}$ & - & - & - & 0.173 & 1.258 & 0.0191 & 0.295\\ 
    & $\mathcal{P}_{\alpha 2}$ & - & - & - & 0.0301 & 0.589 & 0.0133 & 0.730\\
    
    \hline
    \multirow{4}{*}{Fig. 6 (c)} & 
    $\mathcal{P}_c$ & 0.0917 & 0.000204 & 1.817 & - & - & - & -\\ 
    & $\mathcal{P}_b$ & - & - & - & 0.267 & 0.602 & 0.0858 & 0.754\\
    & $\mathcal{P}_{\alpha 1}$ & - & - & - & 0.236 & 2.039 & 0.0181 & 0.322\\ 
    & $\mathcal{P}_{\alpha 2}$ & - & - & - & 0.0309 & 0.448 & 0.0150 & 0.658\\
    
    \hline
    \multirow{4}{*}{Fig. 6 (d)} & 
    $\mathcal{P}_c$ & 0.109 & 0.000214 & 1.652 & - & - & - & -\\ 
    & $\mathcal{P}_b$ & - & - & - & 0.314 & 0.683 & 0.0611 & 0.929\\
    & $\mathcal{P}_{\alpha 1}$ & - & - & - & 0.181 & 1.775 & 0.0221 & 0.318\\ 
    & $\mathcal{P}_{\alpha 2}$ & - & - & - & 0.0787 & 1.061 & 0.00641 & 1.214\\
    
    \hline
    
    \multirow{4}{*}{Fig. 7 (a)} &
    $\mathcal{P}_c$ & 0.102 & 0.000203 & 2.014 & - & - & - & -\\ 
    & $\mathcal{P}_b$ & - & - & - & 2.661 & 4.514 & 0.0324 & 0.618\\
    & $\mathcal{P}_{\alpha 1}$ & - & - & - & 0.213 & 2.437 & 0.0219 & 0.298\\ 
    & $\mathcal{P}_{\alpha 2}$ & - & - & - & 0.0611 & 0.714 & 0.00489 & 1.179\\
    
    \hline
    
    \multirow{4}{*}{Fig. 7 (b)} &
    $\mathcal{P}_c$ & 0.123 & 0.000216 & 1.904 & - & - & - & -\\ 
    & $\mathcal{P}_b$ & - & - & - & 0.532 & 1.042 & 0.0260 & 1.172\\
    & $\mathcal{P}_{\alpha 1}$ & - & - & - & 0.286 & 2.687 & 0.0174 & 0.276\\ 
    & $\mathcal{P}_{\alpha 2}$ & - & - & - & 0.0359 & 0.408 & 0.00941 & 0.728\\
    
    \hline
    
    \multirow{4}{*}{Fig. 7 (c)} &
    $\mathcal{P}_c$ & 0.122 & 0.000207 & 1.962 & - & - & - & -\\ 
    & $\mathcal{P}_b$ & - & - & - & 0.671 & 1.388 & 0.0533 & 1.088\\
    & $\mathcal{P}_{\alpha 1}$ & - & - & - & 0.370 & 2.924 & 0.0123 & 0.293\\ 
    & $\mathcal{P}_{\alpha 2}$ & - & - & - & 0.0350 & 0.418 & 0.0129 & 0.597\\
    
    \hline
    \hline

\end{tabular}
\end{table}

\bibliography{references}{}
\bibliographystyle{aasjournal}

\end{document}